# Polarization-sensitive GeSn Mid-Infrared Membrane Photodetectors with Integrated Plasmonic Metasurface

Ziqiang Cai, Cédric Lemieux-Leduc, Mahmoud R. M. Atalla, Luo Lu, Gérard Daligou, Simone Assali, and Oussama Moutanabbir[#]

*Department of Engineering Physics, École Polytechnique de Montréal, Montréal, C.P. 6079, Succ. Centre-Ville, Montréal, Québec, Canada H3C 3A7*

[#] Corresponding author. Email: oussama.moutanabbir@polymtl.ca

**Abstract**

Germanium-Tin (GeSn) semiconductors are promising for mid-infrared photodetectors owing to their silicon compatibility, tunable bandgap, and potential for room-temperature operation. Released GeSn membranes provide an additional degree of freedom to extend the operation wavelength through epitaxial strain relaxation, while their transferability expands design flexibility. On the other hand, metasurfaces have become an effective strategy to engineer light-matter interaction, and their integration with photodetectors can enhance performance and introduce new functionalities. Here, we demonstrate a mid-infrared photodetector consisting of transfer-printed $Ge_{0.89}Sn_{0.11}$ membrane integrated with a Au plasmonic metasurface. The photodetector exhibits a wavelength cutoff exceeding 3.0 μm and nearly fourfold increase in responsivity at 2.5 μm compared to unreleased films, attributed to Fabry-Pérot resonance. Furthermore, by introducing an anisotropic metasurface, the obtained detectors display a strong polarization sensitivity, achieving a measured contrast ratio of ~4:1 between orthogonal polarizations. Furthermore, the operation wavelength of the photodetector can be selectively tuned by varying the geometric scale of the metasurface. The experimental results show excellent agreement with simulations, confirming the effectiveness and versatility of this integrated metasurface-membrane design.

**Keywords:** Mid-infrared photodetector; Metasurface; Group IV semiconductors; GeSn membrane; Polarization-sensitive detector

**INTRODUCTION**

Mid-infrared (MIR) photodetectors hold strategic importance for sensing and imaging applications, yet their development is hindered by a limited selection of suitable materials.[1] Among the emerging candidates, GeSn alloys have attracted considerable attention as a group-IV platform with tunable optical properties extending into the MIR.[2] Over the past decade, GeSn-based devices have been studied in a variety of room-temperature MIR photodetectors operating between wavelengths of 1.6 and 4 μm,[3-5] including p-i-n photodetectors,[6-13] multiple-quantum-well (MQW) photodetectors,[14-16] waveguide photodetectors,[17, 18] and nanoscale photodetectors.[19, 20] In addition, GeSn alloys have been investigated for other MIR optoelectronic components such as LEDs,[21-25] lasers,[26-34] and modulators.[35] These alloys can be grown epitaxially on Ge/Si substrates, providing compatibility with silicon processing standards.[2] Increasing the Sn composition narrows the bandgap energy of the GeSn alloy. Notably, an indirect-to-direct bandgap transition occurs above 7 at.% Sn in a fully relaxed alloy,[2] significantly improving light absorption and emission. Further Sn incorporation extends the cutoff wavelength deeper into the MIR range.[36] Moreover, the cutoff wavelength of GeSn can also be tuned via engineering its internal strain, by means of controlling the GeSn thickness[37] or etching the underlying sacrificial layer.[28, 38, 39] When the Ge buffer layer is completely removed, the top GeSn layer can be released as a membrane, which relaxes the residual compressive strain and further extends the cutoff wavelength.[36, 40, 41] Furthermore, GeSn membranes offer additional advantages, including reduced dark current due to defects removal,[36] enhanced responsivity,[42, 43] and integration on flexible substrates[44] relevant for printed electronics.[45] While membranes made of semiconductors like Si or Ge have been extensively studied[46-54] and even mass-produced for commercial photonic applications,[55-59] research on GeSn membranes is still in its early stage.

Meanwhile, metasurfaces have emerged as powerful tools for advancing photodetector engineering. As planar, subwavelength arrays of meta-atoms, metasurfaces provide unprecedented control over light by locally manipulating its amplitude, phase, and polarization via careful design for the geometry of each meta-atom, exhibiting enhanced performances and unique functionalities.[60-62] This capability has spurred significant research efforts to harness metasurfaces for enhancing key detector performance metrics, including spectral bandwidth, responsivity, and operational speed.[63, 64] Metasurfaces have been applied in different kinds of photodetectors, such as two-dimensional (2D) material photodetector,[65-67] quantum-dot photodetector,[68, 69] hot-electron photodetector,[70-72] pyroelectric photodetector,[73-75] terahertz photodetector,[76-81], and avalanche photodetector.[82, 83] GeSn photodetectors integrated with metasurfaces have also been subject of recent studies.[84-89] Beyond performance boosting, metasurfaces can also incorporate new functionalities, enabling polarization, multi-dimensional and spectral detections.[64] A photodetector integrated with an asymmetric nanobar array, for example, gains polarization-sensitive responsivity, while combining multiple oriented metasurfaces allows for the measurement of full Stokes parameters.[73, 75, 90-93] Despite this rapid progress in metasurface-integrated photodetectors, the specific combination of a metasurface with a transfer-printed GeSn membrane remains underexplored despite its importance for MIR sensing applications.

In this work, we demonstrate a novel MIR photodetector that integrates transfer-printed

GeSn membranes with anisotropic Au plasmonic metasurfaces. The device was fabricated by first epitaxially growing a GeSn layer (11 at.% Sn) on Si wafer with Ge as interlayer. The $Ge_{0.89}Sn_{0.11}$ layer was then patterned and released as a membrane via selective etching of the buffer layer and transfer-printed onto a $SiO_2$/Si substrate.[40] The high refractive index contrast between the GeSn membrane and the $SiO_2$ substrate enhances Fabry-Pérot (FP) resonances, yielding a nearly fourfold increase in responsivity at the FP resonance (2.5 μm) compared to unreleased films. Moreover, the subsequent integration of Au metasurface makes the photodetector polarization-selective. The responsivity contrast ratio measured at 2.2 μm is approximately 4:1 between orthogonal polarizations. Numerical simulations corroborate the experimental results, showing excellent agreement and validating the proposed design. Finally, we demonstrate that the operating wavelength of the photodetector can be selectively tuned by varying the geometric scale of the metasurface. This result underscores the design flexibility enabled by the integration of free-standing membranes with metasurfaces, providing a versatile platform for wavelength-specific photodetection in the MIR.

**RESULTS AND DISCUSSIONS**

To elucidate the influence of the GeSn membrane structure and the Au metasurface on photodetector performance, we simulated their electric field distributions. Details for the simulation can be found in Methods, and the refractive index of GeSn can be found in Section 1 of Supporting Information. Conventionally, devices are fabricated by epitaxially growing a Ge buffer layer followed by a GeSn layer on a silicon substrate to mitigate defects arising from lattice mismatch (Figure 1a, left panel). Releasing the GeSn membrane by etching the Ge buffer layer and transferring it onto a SiO$_2$/Si substrate (Figure 1a, right panel) serves multiple purposes. First, the released strain extends the photodetector's cutoff wavelength. Second, the high refractive index contrast between GeSn ($n$ = 4.28) and SiO$_2$ ($n$ = 1.44) enables a strong FP resonance within the GeSn layer, thereby enhancing light absorption. Finally, the SiO$_2$ layer acts as an electrically insulating layer for subsequent measurements. As shown in Figure 1b, the distinct fringes in the electric field amplitude distribution within the GeSn membrane confirm the excitation of this FP resonance, which is significantly weaker in the as-grown device. This FP resonance can be further enhanced by replacing the Si substrate with a highly reflective layer, such as a Au layer [11] or a distributed Bragg reflector (DBR).[43] Moreover, the FP resonance can also be enhanced by increasing the thickness of $SiO_2$ layer, or directly using a silica substrate, which can reduce the leakage of light into the substrate.[94]

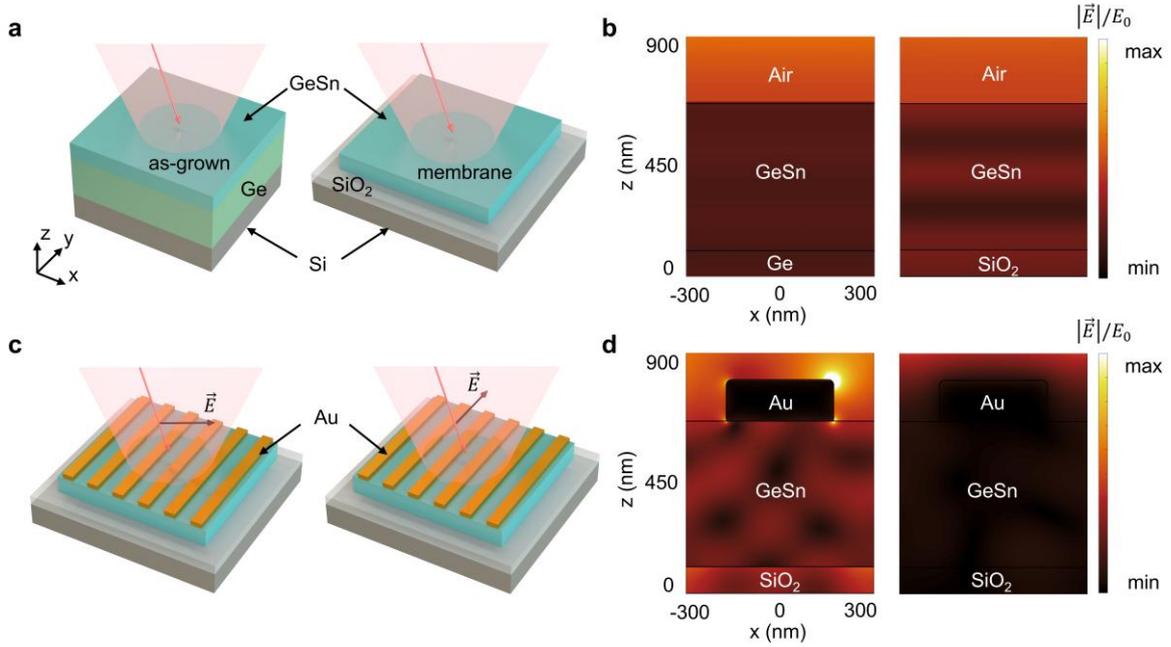

**Figure 1.** Designs of GeSn membrane photodetectors. (a) Schematics of as-grown GeSn photodetector (left) and transferred membrane photodetector (right). (b) Corresponding electric field amplitude distributions of the as-grown (left) and membrane (right) photodetectors. (c) Schematics of membrane photodetector integrated with Au metasurface, with incident electric field to be TM-polarized (left) and TE-polarized (right). (d) Corresponding electric field amplitude distributions for TM-polarized (left) and TE-polarized (right) electric fields, respectively. For (b) and (d), the wavevector of incident light is inside xz-plane with an incident angle of 30 degrees and wavelength of 2.2 μm. $|\vec{E}|/E_0$ refers to the electric field amplitude divided by incident electric field amplitude.

The integration of a metallic metasurface on top of the GeSn membrane photodetector introduces polarization sensitivity to the device. As shown in Figure 1c, a one-dimensional periodic Au grating (150 nm thick, 380 nm width, 710 nm period) was applied to the membrane. This anisotropic structure produces distinct optical responses for orthogonal light polarizations. Figure 1d illustrates the underlying mechanism: under transverse-magnetic (TM) polarized plane wave illumination (electric field polarized in xz-plane), a strong localized enhancement of the electric field occurs due to the excitation of plasmonic resonance (Figure 1d, left panel). In contrast, transverse-electric (TE) polarized light (electric field polarized along y-axis) fails to excite this resonance, resulting in a markedly weaker field intensity (Figure 1d, right panel). This pronounced polarization-dependent response is the foundation for a MIR polarization sensor, with applications in material texture analysis and biomedical imaging.[64, 95]

The device was fabricated according to the process flow illustrated in Figure 2a: (i) Sequential deposition of a 1.02 μm Ge buffer layer and a 520 nm $Ge_{0.89}Sn_{0.11}$ layer on a Si substrate via chemical vapor deposition (CVD); (ii) Patterning of 20 μm × 20 μm mesas via photolithography and wet etching, followed by a dry etching step using $CF_4$ gas to selectively undercut the Ge buffer layer; (iii) Detach of the GeSn membranes from original substrate using a PDMS stamp; (iv) Transfer printing of the membranes onto a $SiO_2$/Si substrate; (v) Post-transfer cleaning and annealing of the membranes; (vi) Fabrication of Au electrodes and the metasurface via a combination of photolithography and electron-beam lithography (EBL).

More details on the GeSn growth and fabrication process can be found in Methods.

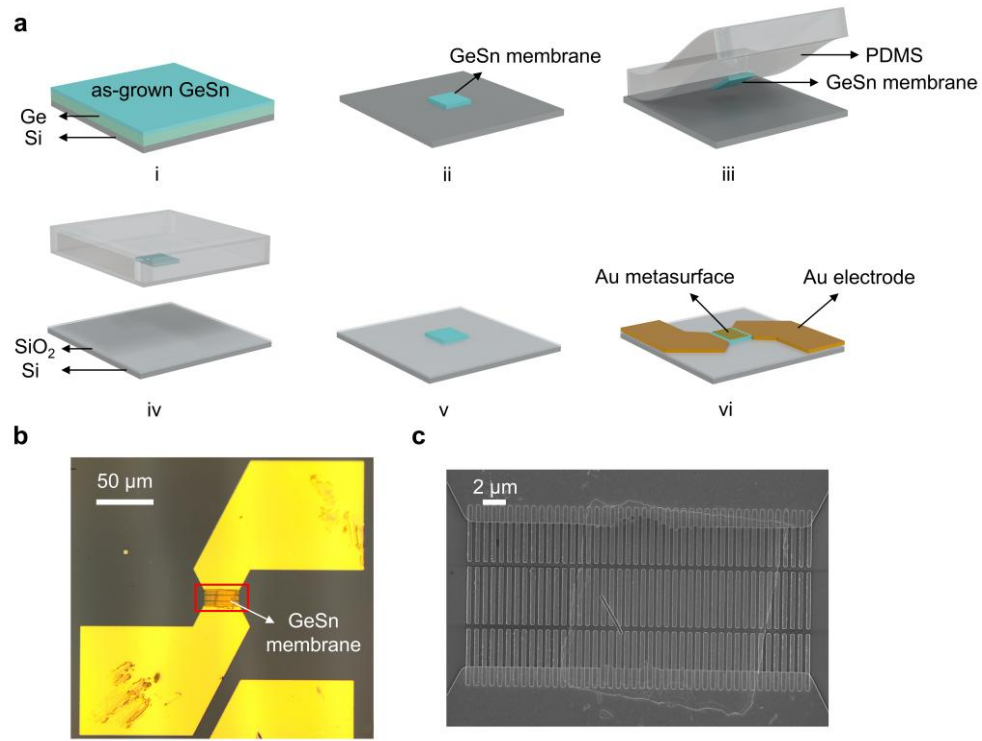

**Figure 2.** (a) Fabrication steps of the GeSn membrane photodetector integrated with Au metasurface. (b) Optical image of the fabricated device. The red box marks the area of SEM image. (c) SEM image of the metasurface on top of the GeSn membrane.

An optical micrograph of the final device is shown in Figure 2b, where the GeSn membrane is contacted by electrodes at both ends and the metasurface is positioned on top. For optoelectrical characterization, a small bias voltage was applied across the electrodes via micromanipulator probes and the resulting current was measured. The incident light generates photocarriers within the membrane, leading to a measurable photocurrent ($I_{photo}$) corresponding to the increase in current above the baseline dark current ($I_{dark}$). The electrical signals were sent to either a Fourier transform spectrometer (FTIR) to retrieve the relative responsivity spectrum, or a source measure unit to measure the absolute responsivity. More details on the FTIR and absolute responsivity measurements can be found in Section 2 of the Supporting Information. Figure 2c shows a scanning electron microscopy (SEM) image of the metasurface region, confirming the fabrication quality of the metasurface and its precise alignment over the GeSn membrane.

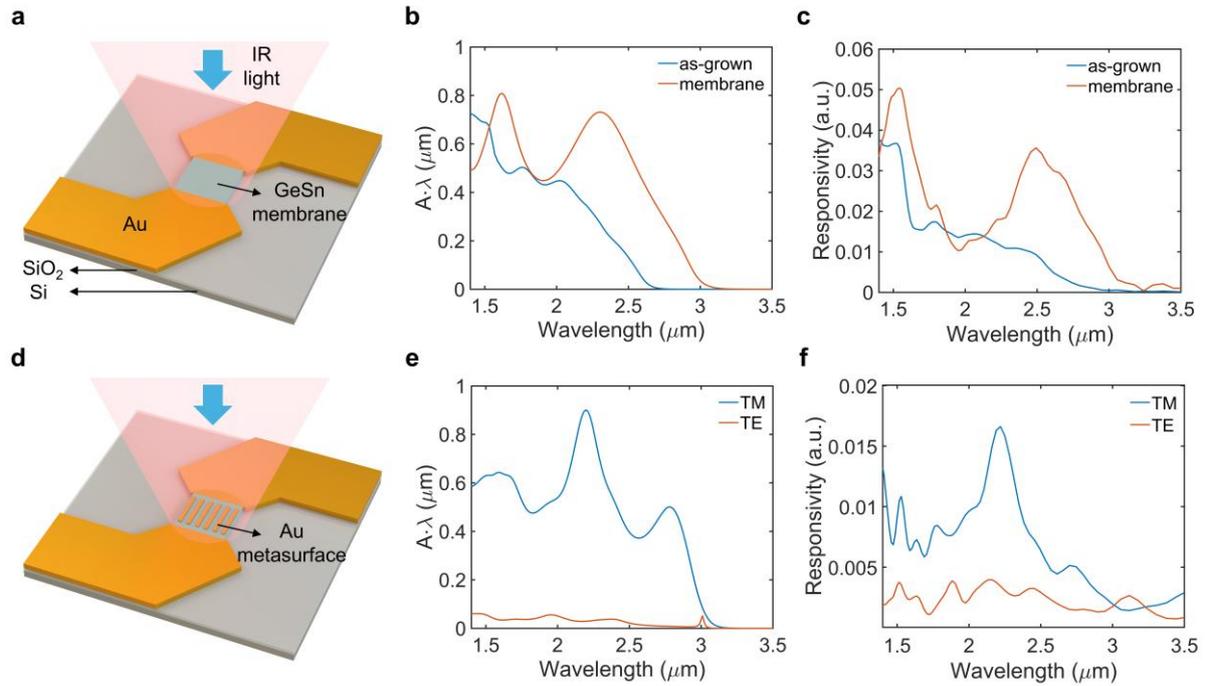

**Figure 3.** Simulated and measured responsivity spectra of the GeSn photodetectors. (a) Schematic of the GeSn membrane photodetector without metasurface. (b) Simulated responsivity spectra for as-grown GeSn photodetector (blue curve) and GeSn membrane photodetector (red curve). (c) Relative responsivity spectra for as-grown photodetector (blue curve) and membrane photodetector (red curve) measured by FTIR (a.u. stands for arbitrary units). (d) Schematic of GeSn membrane photodetector integrated with metasurface. (e) Simulated responsivity spectra for TM-polarized (blue curve) and TE-polarized (red curve) incident light. (f) Relative responsivity spectra for TM-polarized (blue curve) and TE-polarized (red curve) incident light measured by FTIR.

The performance of the GeSn photodetectors was investigated through both numerical simulations and experimental studies. In the simulations (Figure 3b and 3e), the responsivity was estimated from the product of light absorption (A) in GeSn and Ge and wavelength ($\lambda$), based on the principle that responsivity is proportional to the absorbed light intensity divided by the photon energy. Figure 3b illustrates a key distinction between the as-grown and membrane-based devices. For the as-grown photodetector, both the GeSn layer and the underlying Ge buffer contribute to the photoresponse. This results in a characteristic step in the responsivity curve near 1.6 μm, where the Ge contribution begins to decline, diminishing entirely at its cutoff wavelength of 1.8 μm. The GeSn absorption itself falls to zero near its cutoff wavelength of 2.7 μm. In contrast, the GeSn membrane photodetector exhibits a responsivity spectrum with FP resonances. As previously discussed in Figure 1b, these resonances enhance the local optical field within the membrane, thereby boosting the responsivity at resonance wavelengths. Furthermore, strain relaxation associated with the membrane release extends the GeSn cutoff wavelength from 2.7 μm to 3.1 μm. The combined effect of resonant enhancement and a broader spectral range results in a significantly higher responsivity across the 2-3 μm wavelength range. The experimental data, presented in Figure 3c, shows excellent agreement with the simulated results. The measured cutoff wavelength extends from approximately 2.8 μm to 3.2 μm after membrane releasing, and the responsivity exhibits an enhancement of nearly four-fold at 2.5 μm. A minor discrepancy is observed in the membrane device, where the measured resonance peak at longer wavelength is slightly red-shifted compared to the simulation. This small shift is likely due to the discrepancy between the GeSn refractive index used in the model, which was derived from 8-band $k \cdot p$ method calculations,[96,97] and the real refractive index of GeSn. More details on the refractive index used in simulations are provided in Section 1 of the Supporting Information.

To incorporate polarization sensitivity to the device, a Au metasurface was fabricated on the GeSn membrane. The simulated results in Figure 3e reveal a sharp resonance peak at 2.2 μm for TM-polarized light, while the responsivity to TE-polarized light is negligible. This high contrast originates from the polarization-selective nature of the anisotropic structure. The TM-polarized light can efficiently excite plasmonic resonance, leading to strong local field enhancement, whereas TE-polarized light is effectively blocked by the metallic grating. The experimental results in Figure 3f show a resonant peak for TM-polarization and a flat, low response for TE-polarization, agreeing with simulation results. However, the measured polarization contrast is lower than simulated. We attribute this discrepancy to the finite size of the fabricated membrane, considering that our simulations employed periodic boundary conditions that model an infinite array. To validate this explanation, a systematic study showing how the membrane dimensions affect the simulated response is provided in Section 3 of the Supporting Information.

The results shown in Figure 3c and 3f are extracted directly from the FTIR spectrometer and thus are not absolute responsivity values. To quantify the absolute values, we supplemented the relative spectral data from FTIR measurements with absolute responsivities measured at 1.55 μm. This allowed us to deduce the absolute responsivities at the wavelength of interest. At the bias voltage $V_b$ of 30 mV and wavelength $\lambda$ of 2.2 μm, the deduced responsivities for the as-grown and metasurface-integrated membrane photodetectors are 1.52 mA/W and 2.01 mA/W, respectively (details can be found in Section 4 of the Supporting Information). The metasurface-integrated device thus shows an enhancement factor of ~1.3 compared with as-grown GeSn photodetector. This value is lower than the enhancement that GeSn membrane photodetector without metasurface exhibits at 2.5 μm. The greater enhancement of nearly four-fold at 2.5 μm is attributed to its proximity to the cutoff wavelength of as-grown GeSn detector, where the responsivity drops. Moreover, we anticipate that the measured responsivity of metasurface-integrated device can be further enhanced by optimizing the fabrication process to minimize defects, such as membrane edge roughness and metasurface imperfections.

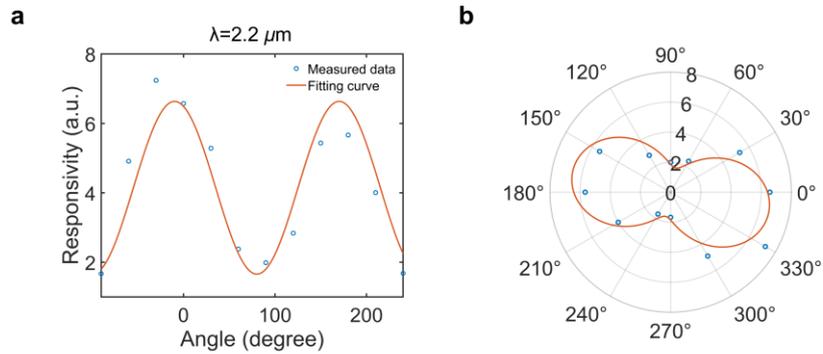

**Figure 4.** (a) Measured responsivity at 2.2 μm versus polarizer angle (blue dots), together with fitting curve (red line). (b) Corresponding polar diagram.

To further investigate the photodetector's polarization sensitivity, we gradually rotate the polarizer and measure the responsivity spectrum at each angle. Corresponding results are presented in Figure 4. The data were fitted with the function $R_x \times \cos^2(\theta + \phi) + R_y \times \sin^2(\theta + \phi)$, where $R_x$ and $R_y$ represent the responsivities for light polarized along the x- and y-axes, respectively. $\theta$ is the polarization angle and $\phi$ is an offset angle. By fitting the data, we retrieve the contrast ratio, defined as $R_x/R_y$, to be 4.0. The original FTIR spectra measured at different angles of polarizer can be found in Section 5 of the Supporting Information.

To explore the geometric tunability of the metasurface, we simulated the photodetector's responsivity spectra across a range of parameters. As shown in Figure 5a, the width and period of the nanoribbons are scaled by the same scale factor (scale), while all other parameters remain constant. In Figure 5b, as scale increases from 0.8 to 1.0, the resonant peak at shorter wavelength redshifts from 2.0 μm to 2.2 μm, and another resonant peak at longer wavelength redshifts from 2.4 μm to 2.8 μm. The resonance wavelength can also be tuned by varying the GeSn layer thickness ($t_{GeSn}$). As shown in Figure 5c and Figure 5d, increasing $t_{GeSn}$ from 450 nm to 550 nm shifts the resonance from 2.1 μm to 2.3 μm. These results confirm that the resonance wavelength can be tuned through either metasurface dimensions or GeSn thickness with high degree of control. However, for practical device fabrication, tuning the metasurface dimensions offers superior advantage compared to modifying the GeSn thickness, since the metasurfaces with different dimensions can be easily fabricated on the same substrate, while thicker GeSn layers are difficult to grow given their metastable nature and large lattice mismatch compared to Ge/Si growth wafers. Furthermore, in our previous work, we have demonstrated transfer printing of an array of GeSn membranes onto a substrate of choice.[40, 41] By fabricating a distinct metasurface on each membrane, and each metasurface is engineered with specific geometric scale to resonate at target wavelength or with defined orientation to resolve individual Stokes parameter, it is possible to develop MIR photodetector for multiwavelength and full-Stokes detection.[92, 93]

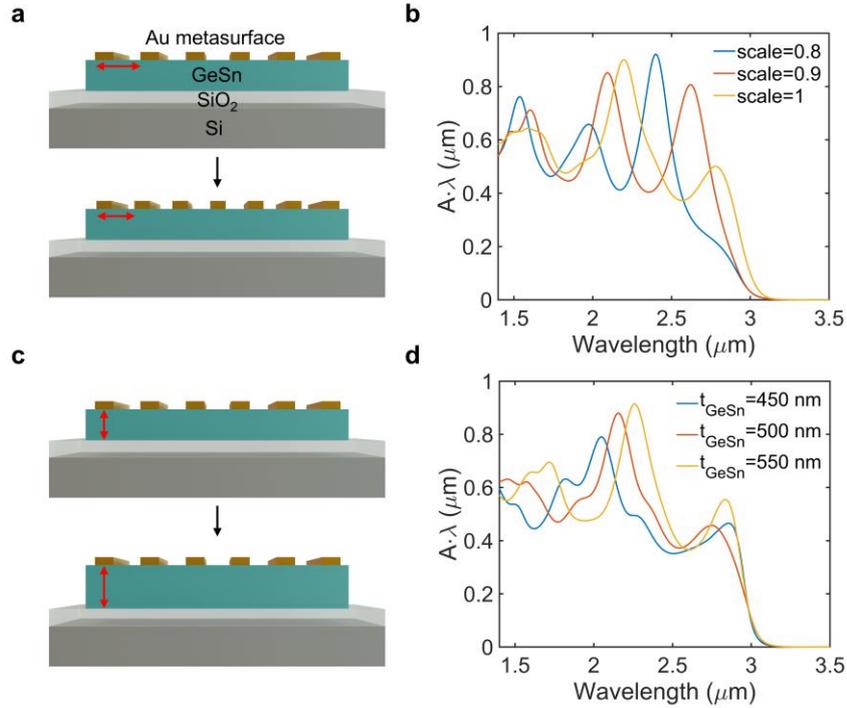

**Figure 5.** (a,b) Geometry and performance of metasurface-tuned photodetectors. (a) Schematic with varying metasurface scales. (b) Corresponding simulation results of responsivity spectra, showing tuning of resonance wavelengths. (c,d) Geometry and performance of thickness-tuned photodetectors. (c) Schematic with varying GeSn layer thicknesses. (d) Corresponding simulation results. The red arrows in (a) and (c) mark the changes in geometry.

**Conclusion**

We have demonstrated a polarization-sensitive MIR photodetector by integrating a Au plasmonic metasurface onto transfer-printed $Ge_{0.89}Sn_{0.11}$ membrane. Releasing the membrane by etching the underlying Ge layer redshifts the cutoff wavelength from 2.7 μm to 3.1 μm. When transferred to a low refractive index substrate, the integrated detector exhibits FP resonances that enhance the responsivity in the 2-3 μm wavelength range, with nearly four-fold enhancement at 2.5 μm as compared to as-grown devices. The metasurface-integrated photodetector displays strong polarization sensitivity, with a measured contrast ratio of 4.0 at 2.2 μm. Furthermore, by modifying the metasurface geometry, the resonance wavelength can be tuned within the 2.0-2.8 μm range. This approach can broaden the application of GeSn photodetectors, paving the way for high performance room temperature multifunctional MIR photodetectors.

**Acknowledgements**

The authors acknowledge support from NSERC Canada, Canada Research Chairs, Canada Foundation for Innovation, Mitacs, PRIMA Québec, and Defense Canada (Innovation for Defense Excellence and Security, IDEaS), the European Union's Horizon Europe research and innovation program under grant agreement No 101070700 (MIRAQLS), and the US Air Force Research Office Grant No. W911NF-22-1-0277.

**Methods**

*Numerical simulation:* We performed numerical simulations using the Electromagnetic Waves Frequency Domain module in COMSOL Multiphysics. A 2D model with periodic boundary conditions was employed to simplify the calculations. Linearly-polarized plane-waves with either TM or TE polarizations were applied as incident lights. The angle of incidence is 30 degree to match the Cassegrain lens configuration of our experimental setup. Absorptions of light in GeSn and Ge were calculated by integrating the electromagnetic power loss density over the GeSn and Ge area, then divided by the input power.

*Epitaxial growth of GeSn layers:* The $Ge_{0.89}Sn_{0.11}$ alloy growth was carried out in a low-pressure chemical vapor deposition (LPCVD) reactor on a 4-inch Si (100) wafer. First, a 1.02 μm-thick Ge virtual substrate (Ge-VS) was grown using 10% monogermane ($GeH_4$) and a continuous flow of pure hydrogen following a two-temperature growth protocol at 460 ºC and 600 ºC. In addition to reducing the lattice mismatch between the Si substrate and GeSn, Ge-VS also acts as a sacrificial layer, facilitating the GeSn membranes under-etching and release. Subsequently, the $Ge_{0.89}Sn_{0.11}$ layers were grown on Ge/Si wafers at 320 ºC using $GeH_4$ and $SnCl_4$ precursors. The microstructure of the as-grown material was assessed using cross-sectional transmission electron microscopy and the Sn composition was determined through X-ray diffraction with reciprocal space mapping.

*Fabrication:* The device was fabricated following the process flow in Figure 2a. For membrane fabrication, after the epitaxial growth of GeSn layer, we first diced the wafer. Then we spin-coated photoresist (PR) onto the chip and used photolithography to define the patterns of GeSn membranes, followed by a post-bake process to enhance the adhesion between PR and GeSn layer. Next, the patterned PR was used as etch masks, and the unprotected areas were etched down by Cr etchant solution. After removing the PR with acetone, the remaining Ge layer was removed using reactive ion etching (RIE) with a $CF_4$ gas. After GeSn membranes were fabricated, we used a PDMS stamp coated with polycarbonate to catch the membranes. Following that, the PDMS with the membranes were pressed onto a new substrate, which was prepared by depositing 100 nm $SiO_2$ layer on a silicon wafer using plasma-enhanced chemical vapor deposition (PECVD). Then the substrate was heated to 180 ºC to melt the polycarbonate, and the PDMS stamp was lifted with membranes remaining on the substrate, followed by a cleaning process to remove the polycarbonate on the membranes. Afterwards, we used another round of photolithography to pattern the electrodes. We deposit 5 nm Ti and 150 nm Au by e-beam evaporation and lift-off process to fabricate the electrodes. Finally, we used EBL process to define the metasurface patterns, followed by another round of e-beam evaporation and lift-off process to fabricate the metasurfaces onto the GeSn membranes.

# Supporting Information for
# Polarization-sensitive GeSn Mid-Infrared Membrane Photodetectors with Integrated Plasmonic Metasurface

Ziqiang Cai, Cédric Lemieux-Leduc, Mahmoud R. M. Atalla, Luo Lu, Gérard Daligou, Simone Assali, and Oussama Moutanabbir[#]

*Department of Engineering Physics, École Polytechnique de Montréal, Montréal, C.P. 6079, Succ. Centre-Ville, Montréal, Québec, Canada H3C 3A7*
[#] Corresponding author. Email: oussama.moutanabbir@polymtl.ca

## 1. Refractive Index of GeSn

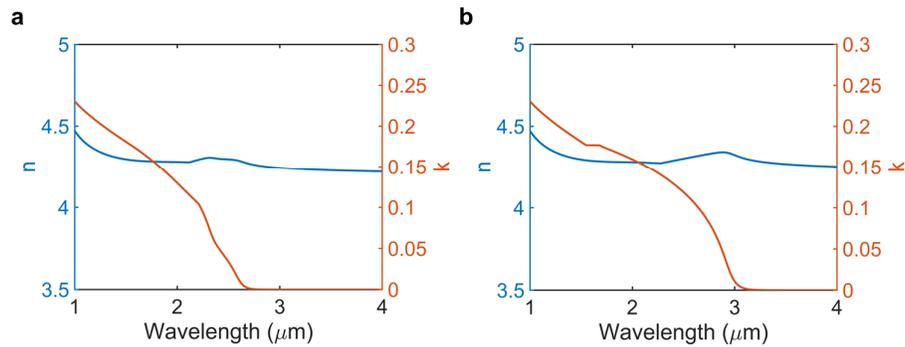

**Figure S1.** Real part n and imaginary part k of the refractive index of GeSn before (a) and after (b) releasing.

In the as-grown samples, the GeSn layer contains strain due to the lattice mismatch between layers during the chemical vapor deposition (CVD) growth. This strain is released after the underlying material is etched, creating a GeSn membrane. The resulting change in internal strain alters the band structure, leading to a substantial shift in the GeSn layer's cut-off wavelength. Accurate simulations require precise refractive index data for each material. However, the small size of the GeSn membrane makes it impossible to directly measure the refractive index via ellipsometry. Consequently, we can only measure the refractive index of the as-grown GeSn sample, which we have done over a wavelength range from 0.21 μm to 2.5 μm. The refractive index for longer wavelengths must be calculated.

The refractive index of GeSn was figured out by first calculating the absorption coefficient using the 8-band $k \cdot p$ method [1,2], and then applying the Kramers-Krönig relation [3] to get the real part. The composite refractive index profiles used in our simulations are shown in Figure S1. For both as-grown GeSn and released membrane, the profile is a combination of directly measured values at shorter wavelengths and calculated values at longer wavelengths. The redshift in the cutoff wavelength from 2.7 μm to 3.1 μm after strain release, as shown in Figure S1a and S1b, agrees well with the experiment data.

2. **Measurement Setups**

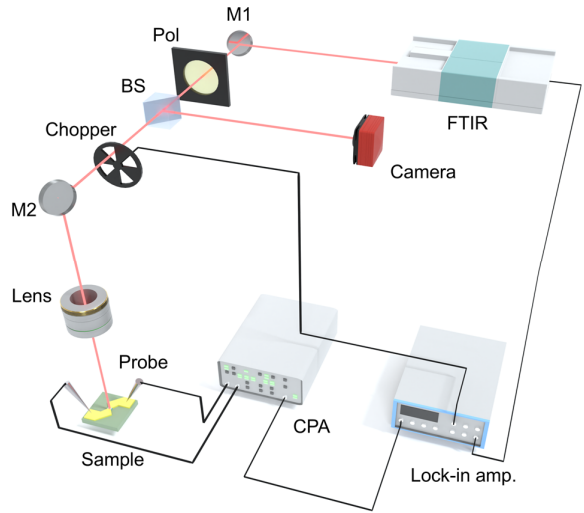

**Figure S2.** Schematic of the optical setup for the FTIR spectrum. 'M1' and 'M2' represent reflective mirrors. 'Pol' and 'BS' refer to polarizer and beam-splitter, respectively. CPA stands for current pre-amplifier. The red lines represent infrared light beams, and the black lines represent the electric wires.

The photodetector performance was characterized by the optical setup illustrated in Figure S2. Infrared light from a Fourier-transform infrared (FTIR) spectrometer (Bruker Vertex 70) was passed through a polarizer, a beam splitter, and a chopper before being focused onto the sample by a Cassegrain objective lens (Thorlabs LMM40X-UVV). The generated photocurrent was collected by electrical probes, then amplified sequentially by a current pre-amplifier (CPA) and a lock-in amplifier. The resulting signal was returned to the FTIR for processing. The use of a chopper, synchronized with the CPA and lock-in amplifier, significantly enhanced the signal-to-noise ratio. The FTIR spectrum was obtained by recording the signal while varying the optical path length difference in the interferometer, followed by a Fourier transformation to the recorded interferogram. Finally, the FTIR spectrum was normalized by the power spectrum of the light source of FTIR.

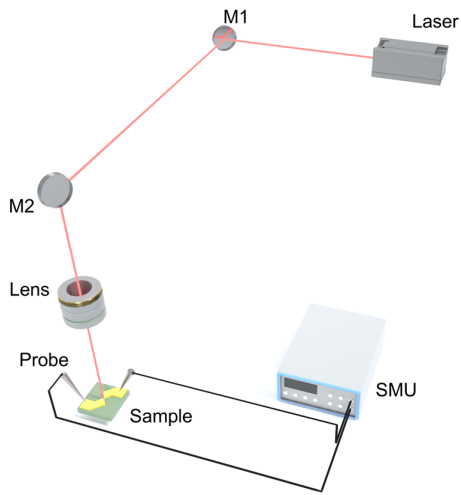

**Figure S3.** Schematic of the optical setup for the responsivity measurement. 'SMU' stands for source measure unit.

The absolute responsivity of the photodetector was measured using the setup depicted in Figure S3. The FTIR was replaced with a 1.55 μm infrared laser. The laser beam, with an optical power of 1.26 mW, was focused onto the GeSn membrane with a spot size of 7 μm in diameter. The photocurrent and dark current were measured using a source measure unit (SMU, Keithley 2470). To ensure the laser was optimally focused on the membrane, the sample stage was carefully adjusted until the photoresponse was maximized. Subsequently, current–voltage (I–

V) characteristics were recorded with or without laser illumination to measure photo and dark currents. The photocurrent was calculated by subtracting the dark current from the total current measured under laser illumination.

## 3. Simulations of finite-sized GeSn membranes

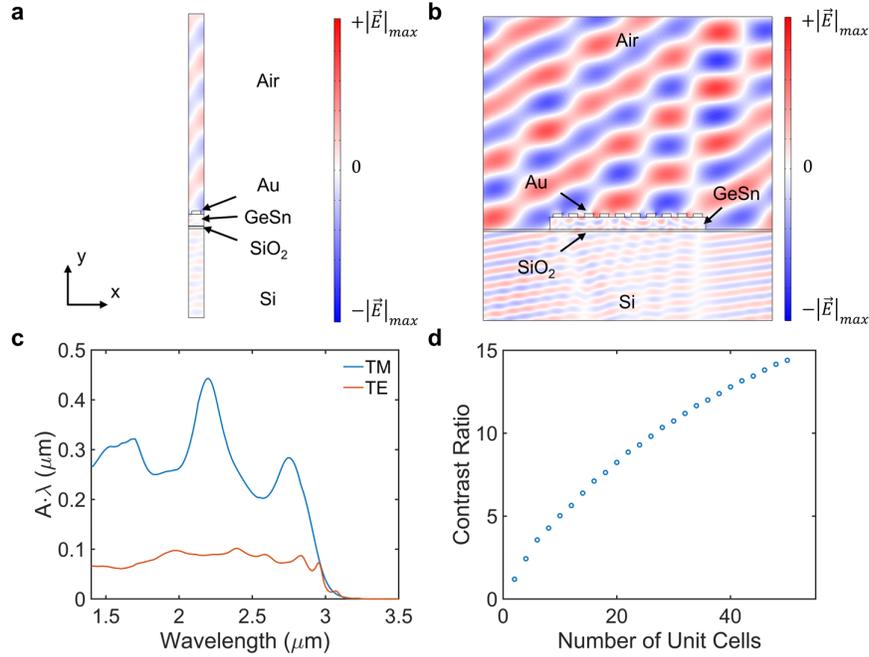

**Figure S4.** (a) Electric field distribution in one unit cell with periodic boundary conditions. (b) Electric field distribution for finite-sized membrane (10 unit cells) with perfect matching layer (PML) boundary conditions. (c) Corresponding responsivity spectra for finite-sized membrane. (d) Contrast ratios at 2.2 μm for finite-sized membranes with different number of unit cells. Incident wave is TM-polarized with a 30-degree angle of incidence.

Simulations of periodic metasurfaces typically employ periodic boundary conditions, which model an infinite array by calculating only a single unit cell. However, all metasurfaces are finite, leading to discrepancies between simulated and measured results. These discrepancies are particularly significant for smaller devices. The active area of our GeSn membrane photodetector is approximately 18 μm by 12 μm, which is not large relative to the operating wavelengths. To investigate the impact of finite size on device performance, we simulated the optical response of a finite-sized membrane (Figure S4b and S4c), which has 10 unit cells as an example. For comparison, Figure S4a present the simulated electric field distribution under periodic boundary conditions, and corresponding responsivity spectrum is shown in Figure 3e of the main text. Comparing these results, it reveals that the contrast ratio is obviously lower for the finite-sized membrane. This explains why the measured contrast ratio is smaller than the initial simulation result. Furthermore, the electric field distribution in Figure S4b shows distinct diffraction patterns at the membrane edges compared to the center, illustrating the edge effect that alters the optical performance of device. We also simulated the contrast ratio for different membrane sizes, shown in Figure S4d. As the number of unit cells increases, the contrast ratio rises and will eventually saturate at the value of 31 for an infinitely large membrane. Note that we used 2D simulations for simplicity and didn't consider the finite dimension along the out-of-plane axis, which could further reduce the contrast ratio.

4. **Measured Absolute Resoponsivities**

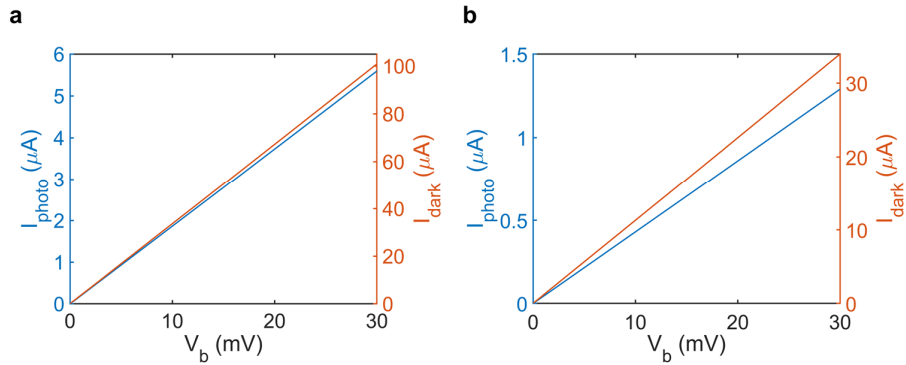

**Figure S5.** (a) Measured photocurrent ($I_{photo}$) and dark current ($I_{dark}$) of as-grown GeSn photodetector. (b) Measured $I_{photo}$ and $I_{dark}$ of GeSn membrane photodetector integrated with metasurface. The measurements are all taken at wavelength of 1.55 μm.

Figure S5 presents the measured $I_{photo}$ and $I_{dark}$ of the as-grown GeSn photodetector (a) and the GeSn membrane photodetector integrated with metasurface (b), using the measurement setup shown in Figure S3. Corresponding responsivities can be calculated through dividing $I_{photo}$ by the optical power (1.26 mW) of incident laser beam. At a bias of 30 mV and wavelength of 1.55 μm, the as-grown device exhibits a responsivity of 4.43 mA/W, compared to 1.02 mA/W for the metasurface device. However, at a longer wavelength of 2.2 μm (also at 30 mV bias), the responsivities are 1.52 mA/W for the as-grown device and 2.01 mA/W for the metasurface device, deduced from FTIR spectra in Figure 3c and Figure 3f of the main text.

## 5. FTIR Spectra at Different Polarizer Angles

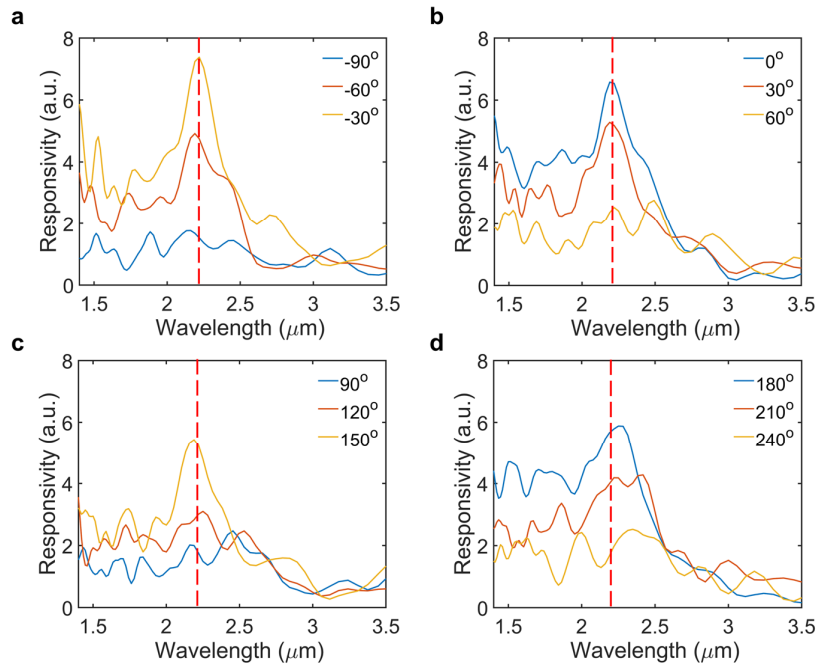

**Figure S6.** Responsivity spectra measured by FTIR with different polarizer angles. Red dashed lines indicate the wavelength of interest at 2.2 μm.

Figure S6 presents the responsivity spectra measured by FTIR with different polarizer angles. The data shown in Figure 4a is obtained by fixing the wavelength at 2.2 μm.